\documentstyle[12pt,bezier]{article}
 
\makeatletter
 
\def\section{\@startsection {section}{1}{\z@}{+3.0ex plus +1ex minus
  +.2ex}{2.3ex plus .2ex}{\normalsize\bf}}
 
\expandafter\ifx\csname mathrm\endcsname\relax\def\mathrm#1{{\rm #1}}\fi
 
\makeatletter

\newcount\@tempcntc
\def\@citex[#1]#2{\if@filesw\immediate\write\@auxout{\string\citation{#2}}\fi
  \@tempcnta\z@\@tempcntb\m@ne\def\@citea{}\@cite{\@for\@citeb:=#2\do
    {\@ifundefined
       {b@\@citeb}{\@citeo\@tempcntb\m@ne\@citea
        \def\@citea{,\penalty\@m\ }{\bf ?}\@warning
       {Citation `\@citeb' on page \thepage \space undefined}}%
    {\setbox\z@\hbox{\global\@tempcntc0\csname
b@\@citeb\endcsname\relax}%
     \ifnum\@tempcntc=\z@ \@citeo\@tempcntb\m@ne
       \@citea\def\@citea{,\penalty\@m}
       \hbox{\csname b@\@citeb\endcsname}%
     \else
      \advance\@tempcntb\@ne
      \ifnum\@tempcntb=\@tempcntc
      \else\advance\@tempcntb\m@ne\@citeo
      \@tempcnta\@tempcntc\@tempcntb\@tempcntc\fi\fi}}\@citeo}{#1}}

\def\@citeo{\ifnum\@tempcnta>\@tempcntb\else\@citea
  \def\@citea{,\penalty\@m}%
  \ifnum\@tempcnta=\@tempcntb\the\@tempcnta\else
   {\advance\@tempcnta\@ne\ifnum\@tempcnta=\@tempcntb \else
\def\@citea{--}\fi
    \advance\@tempcnta\m@ne\the\@tempcnta\@citea\the\@tempcntb}\fi\fi}

\def\nl{\nonumber\\}

\def\beq{\begin{equation}}
\def\eeq{\end{equation}}
\def\beqar{\begin{eqnarray}}
\def\eeqar{\end{eqnarray}}
\def\barr#1{\begin{array}{#1}}
\def\earr{\end{array}}
\def\bfi{\begin{figure}}
\def\efi{\end{figure}}
\def\btab{\begin{table}}
\def\etab{\end{table}}
\def\bce{\begin{center}}
\def\ece{\end{center}}

\def\text{\textstyle}

\def\al{\alpha}
\def\be{\beta}
\def\Ga{\Gamma}
\def\ga{\gamma}
\def\de{\delta}

\def\veps{\varepsilon}
\def\la{\lambda}

\def\si{\sigma}
\def\Si{\Sigma}

\def\refeq#1{\mbox{(\ref{#1})}}
\def\refeqs#1{\mbox{(\ref{#1})}}
\def\refeqf#1{\mbox{(\ref{#1})}}
\def\reffi#1{\mbox{Fig.~\ref{#1}}}

\def\citere#1{\mbox{Ref.~\cite{#1}}}
\def\citeres#1{\mbox{Refs.~\cite{#1}}}
 
\renewcommand{\L}{{\cal{L}}}
 
\def\mathswitchr#1{\relax\ifmmode{\mathrm{#1}}\else$\mathrm{#1}$\fi}
\newcommand{\Pf}{\mathswitch  f}
\newcommand{\Pfbar}{\mathswitch{\bar f}}
\newcommand{\PW}{\mathswitchr W}
\newcommand{\PZ}{\mathswitchr Z}
\newcommand{\PA}{\mathswitchr A}

\newcommand{\PH}{\mathswitchr H}

\newcommand{\Pt}{\mathswitchr t}
\newcommand{\Ptbar}{\mathswitchr{\bar t}}

\newcommand{\PWp}{\mathswitchr {W^+}}
\newcommand{\PWm}{\mathswitchr {W^-}}
 
\def\mathswitch#1{\relax\ifmmode#1\else$#1$\fi}

\newcommand{\MW}{\mathswitch {M_\PW}}

\newcommand{\MZ}{\mathswitch {M_\PZ}}

\newcommand{\sw}{\mathswitch {s_\PW}}
\newcommand{\cw}{\mathswitch {c_\PW}}

\newcommand{\Qf}{\mathswitch {Q_\Pf}}

\newcommand{\vf}{\mathswitch {v_\Pf}}
\newcommand{\af}{\mathswitch {a_\Pf}}

\def\d{{\mathrm{d}}}
\def\ie{i.e.\ }
\def\eg{e.g.\ }

\newcommand{\se}{self-energy}
\newcommand{\ses}{self-energies}

\newcommand{\mpar}[1]{{\marginpar{\hbadness10000%
                      \sloppy\hfuzz10pt\boldmath\bf#1}}%
                      \typeout{marginpar: #1}\ignorespaces}
\def\draftdate{\relax}
\def\mda{\relax}
\def\mua{\relax}
\def\mla{\relax}
\def\draft{
\def\thtystars{******************************}
\def\sixtystars{\thtystars\thtystars}
\typeout{}
\typeout{\sixtystars**}
\typeout{* Draft mode!
         For final version remove \protect\draft\space in source file *}
\typeout{\sixtystars**}
\typeout{}
\def\draftdate{\today}
\def\mda{\mpar{$\downarrow$}}
\def\mua{\mpar{$\uparrow$}}
\def\mla{\marginpar[\boldmath\hfil$\rightarrow$\hfil]%
                   {\boldmath\hfil$\leftarrow $\hfil}%
                    \typeout{marginpar:
$\leftrightarrow$}\ignorespaces}
\def\mua{\marginpar[\boldmath\hfil$\uparrow$]%
                   {\boldmath$\uparrow$\hfil}%
                    \typeout{marginpar: $\uparrow$}\ignorespaces}
\def\mda{\marginpar[\boldmath\hfil$\downarrow$]%
                   {\boldmath$\downarrow$\hfil}%
                    \typeout{marginpar: $\downarrow$}\ignorespaces}
\def\mla{\marginpar[\boldmath\hfil$\rightarrow$]%
                   {\boldmath$\leftarrow $\hfil}%
                    \typeout{marginpar: $\leftrightarrow$}\ignorespaces}
\overfullrule 5pt
\oddsidemargin -15mm
\marginparwidth 29mm
}
 
\def\stars{\strut\leaders\hbox{*}\hfill\strut}
\def\starline{\hfil\strut\hfil\hbox to \textwidth {\stars}\hfil}

\def\pt{pinch-technique}
\def\pt{PT}
\def\bfm{BFM}
\def\Vhat{\hat V}
\def\xiQ{\xi_Q}
\def\xiA{\hat\xi}

\begin{document}
 
\thispagestyle{empty}
\def\thefootnote{\fnsymbol{footnote}}
\setcounter{footnote}{1}
\null
\draftdate \hfill BI-TP. 94/17
\vskip 1cm
\vfill
\begin{center}
{\Large \bf 
\boldmath{Gauge Invariance of Green Functions: 
Background-Field Method versus Pinch Technique}
\par} \vskip 2.5em
{\large
{\sc A.~Denner, G.~Weiglein} \\[1ex]
{\normalsize \it Institut f\"ur Theoretische Physik, Universit\"at W\"urzburg\\
Am Hubland, D-97074 W\"urzburg, Germany}
\\[2ex]
{\sc S.~Dittmaier%
\footnote{Supported by the Bundesministerium f\"ur Forschung und
Technologie, Bonn, Germany.} \\[1ex]
{\normalsize \it Theoretische Physik, Universit\"at Bielefeld\\ 
Universit\"atsstra{\ss}e, D-33501 Bielefeld, Germany}
}
\par} \vskip 1em
\end{center} \par
\vskip 2cm 
\vfil
{\bf Abstract:} \par
Application of the background-field method to QCD and
the electroweak Standard Model yields gauge-invariant effective 
actions giving rise to simple Ward identities.
Within this method, we calculate the quantities 
that have been treated in the literature using the pinch technique. 
Putting the quantum gauge parameter equal to one, we recover the 
pinch-technique results as a special case of the background-field
method.
The one-particle-irreducible Green functions
of the background-field method fulfil for arbitrary gauge parameters
the desirable theoretical properties
that have been noticed within the pinch technique.
Therefore the background-field formalism provides
a general framework for the direct calculation of well-behaved Green
functions.
Within this formalism,
the pinch technique appears as one of arbitrarily many
equivalent possibilities.

\par
\vskip 1cm 
\noindent BI-TP. 94/17 \par
\vskip .15mm
\noindent May 1994 \par
\null
\setcounter{page}{0}
\clearpage
\def\thefootnote{\arabic{footnote}}
\setcounter{footnote}{0}
 
All known successful theories describing the interactions of elementary
particles are gauge theories. However, in order to evaluate quantized
gauge theories within perturbation theory, one has to break gauge
invariance in intermediate steps by choosing a definite gauge. 
As a consequence, although the 
physical observables, \ie the S-matrix elements, are 
gauge-independent, the Green functions, the building blocks of the 
S-matrix elements, are gauge-dependent in the conventional formalism. 

Before we proceed, we remind the reader of the notion of gauge invariance
and gauge independence: gauge invariance means  invariance under gauge
transformations. 
The gauge invariance of the classical Lagrangian gives rise to Ward
identities between the Green functions of the quantized theory.
Gauge independence becomes relevant when quantization
is done by fixing  a gauge. It means independence of the method of gauge
fixing. 

The gauge dependence of Green functions poses no problem as long
as one calculates physical observables in a fixed order of perturbation
theory. 
However, as soon as one does not take into account all
contributions in a given order one will in general arrive at 
gauge-dependent results. 
This happens usually if one tries to resum higher-order corrections
via Dyson summation of \ses\ or if one is only interested in
particular contributions like definite formfactors, \eg magnetic moments
for off-shell particles, without taking into account the full set of diagrams.
This has been frequently done in the literature.

Motivated by these facts, various attempts have been made 
to define gauge-indepen\-dent building blocks.
In order to  construct gauge-independent running couplings, several
proposals for gauge-independent \ses\ have been made
\cite{Ke89,Ku91}. These were essentially obtained by considering
four-fermion processes and shifting parts of the box and vertex diagrams
to the \ses\ to cancel the gauge-parameter dependence 
of the latter within the class of $R_\xi$ gauges. As one can shift
arbitrary gauge-independent contributions between the different building
blocks the resulting quantities are not unique. This freedom has been 
used to require certain desirable properties from the \ses,
like a decent asymptotic behaviour and the vanishing of the
photon--\PZ-boson mixing at zero momentum transfer.
It nevertheless resulted in different definitions of gauge-independent 
building blocks.
All these ad-hoc treatments only refer to four-fermion processes
and do not give a general prescription which is applicable to other
vertex functions.

Such a prescription is given by the so-called pinch technique (\pt)
\cite{Co82,Co89,Pa90,De92}. The \pt\ is an algorithm for the 
construction of  (within $R_\xi$ gauges) gauge-independent vertex functions by
reorganizing parts of the Feynman diagrams contributing to a manifestly
gauge-independent quantity, leaving only a trivial gauge dependence in the
tree propagators. The results obtained via the \pt\ 
directly fulfil the desirable
properties that had to be explicitly enforced in the ad-hoc treatments
mentioned above. But even more important, it turns out that the 
vertex functions constructed according to the \pt\
fulfil the simple Ward identities related to the
classical Lagrangian.

However, the \pt\ leaves many
questions unanswered. So far, it has only been realized for specific
vertex functions at the one-loop level. 
Its application to other vertex functions is not always clear and
its generalization to higher orders is non-trivial and non-unique.
Although the \pt\ vertex functions are claimed to be process-%
independent this has to the best of our knowledge not been proven but
only shown for specific examples. 
It is very unsatisfactory that 
no explanation exists for the fact that the 
\pt\ rules yield vertex functions with the desirable properties and
in particular that these vertex functions fulfil simple Ward identities.
Finally, although the application of the \pt\ to four-fermion processes
is rather simple, it turns out that
the explicit construction of general \pt\ vertex
functions can be technically quite involved.

The purpose of this letter is to provide some insight into these 
problems.
We will show that the existing results
obtained via the \pt\ can be reproduced  as a special case within the 
background-field method (\bfm). 
The \bfm\ provides a natural explanation for the desirable properties of
the vertex functions noticed in the \pt\ by relating them to the 
Ward identities of the \bfm\ which follow directly from gauge invariance.
The \bfm\ is applicable to all orders  of perturbation theory 
and for all vertex functions.
It not only generalizes the \pt\ but in addition yields 
an infinite set of different vertex functions fulfilling the same
properties.
The vertex functions of the \bfm\ are directly derived from
Feynman rules. Since no complicated rearrangement between
different Green functions is needed, the \bfm\ is technically much
simpler than the \pt\ and the vertex functions obviously are not
plagued with any process dependence.
\bigskip

The \bfm\ \cite{Ab81} is a technique for 
quantizing gauge theories without losing explicit gauge invariance. 
To this end one decomposes in the classical Lagrangian
the usual gauge field $V'$ into a 
background field $\Vhat$ and a quantum field $V$ 
and adds a suitable non-linear 
gauge-fixing term such that a gauge-invariant effective action 
$\Ga[\Vhat]$ can be constructed.
Its invariance with respect to background-field 
gauge transformations gives rise to
simple Ward identities between the corresponding vertex functions
which follow from
\beq
\de_{\mathrm{gauge}}\; \Gamma[\Vhat] = 0.
\eeq
The S-matrix is constructed in the usual way by
forming trees with vertices from $\Gamma[\Vhat]$ connected by
lowest-order background-field propagators \cite{Ab83}.

The background-field vertex functions can be calculated using Feynman 
rules that distinguish between quantum fields and background fields. 
Whereas the quantum fields appear only inside loops the background
fields appear only in tree lines. The gauge fixing of the background
and quantum fields is completely unrelated resulting in independent
gauge-parameters $\xiA$ and $\xiQ$, respectively.
The gauge-invariant effective action depends only on the quantum gauge
parameter $\xiQ$. The background gauge parameter $\xiA$ only enters the 
S-matrix elements via the tree propagators.
There are no background ghost fields and 
there is in general no need to split the matter fields. Thus, fermion and scalar
fields can be treated as usual, they have the conventional Feynman rules,
and there is no distinction between background and quantum fields.
\bigskip

\def\GVVV{\bar\Ga}
The relation between the \bfm\ and the \pt\ can be most easily seen in QCD.
The Feynman rules for QCD within the \bfm\ have
been given in \citere{Ab81}. They are particularly simple in the 
Feynman gauge, $\xiQ=1$.  
For example, the coupling of one background gluon with momentum $q^\mu$
to two quantum gluons reads
\beq
\Ga^{abc}_{\al\mu\la}(k_1,q,k_2) =
-g f^{abc} \GVVV_{\al\mu\la}(k_1,q,k_2)  = -g f^{abc} [g_{\al\la} (k_1 - k_2)_\mu -
2 g_{\mu\la} q_{\al} + 2 g_{\mu\al} q_{\la}],
\eeq
where $f^{abc}$ are the structure constants of the gauge group $SU(N)$
and all momenta are incoming.
Straightforward application of those rules for $\xiQ=1$
yields for the one-loop gluon \se\ within dimensional regularization 
(we omit the fermion contribution) 
\beqar \label{gg}
\Si_{\mu\nu}^{ab}(q) &=& -i\Ga_{\mu\nu}^{ab,(1)}(q)  \nl[1ex]
&=& -\de^{ab}\frac{i g^2 N}{(2\pi)^D} 
\mu^{4-D} \int\frac{\d^D k}{k^2(q+k)^2}
\biggl[\frac{1}{2}\GVVV^{\al}_{\phantom{\al}\mu\la}(k,q,-q-k)
\GVVV_{\al\nu}^{\phantom{\al\nu}\la}(k,q,-q-k) \nl[1ex]
&& \qquad {}- (2k+q)_\mu (2k+q)_\nu\biggr],
\eeqar
where the first term originates from the gluon loop (\reffi{FIgg}a)
and the second term from the ghost loop (\reffi{FIgg}b).
The graphs involving quartic couplings vanish within dimensional
regularization.
The result \refeq{gg} is precisely the same as the one derived within the \pt\ 
as given in (3.9) of \citere{Co89}. 
\bfi
\begin{picture}(350,60)
\put(30,50){a)}
\put(170,50){b)}
\put(-40,-465){\includegraphics{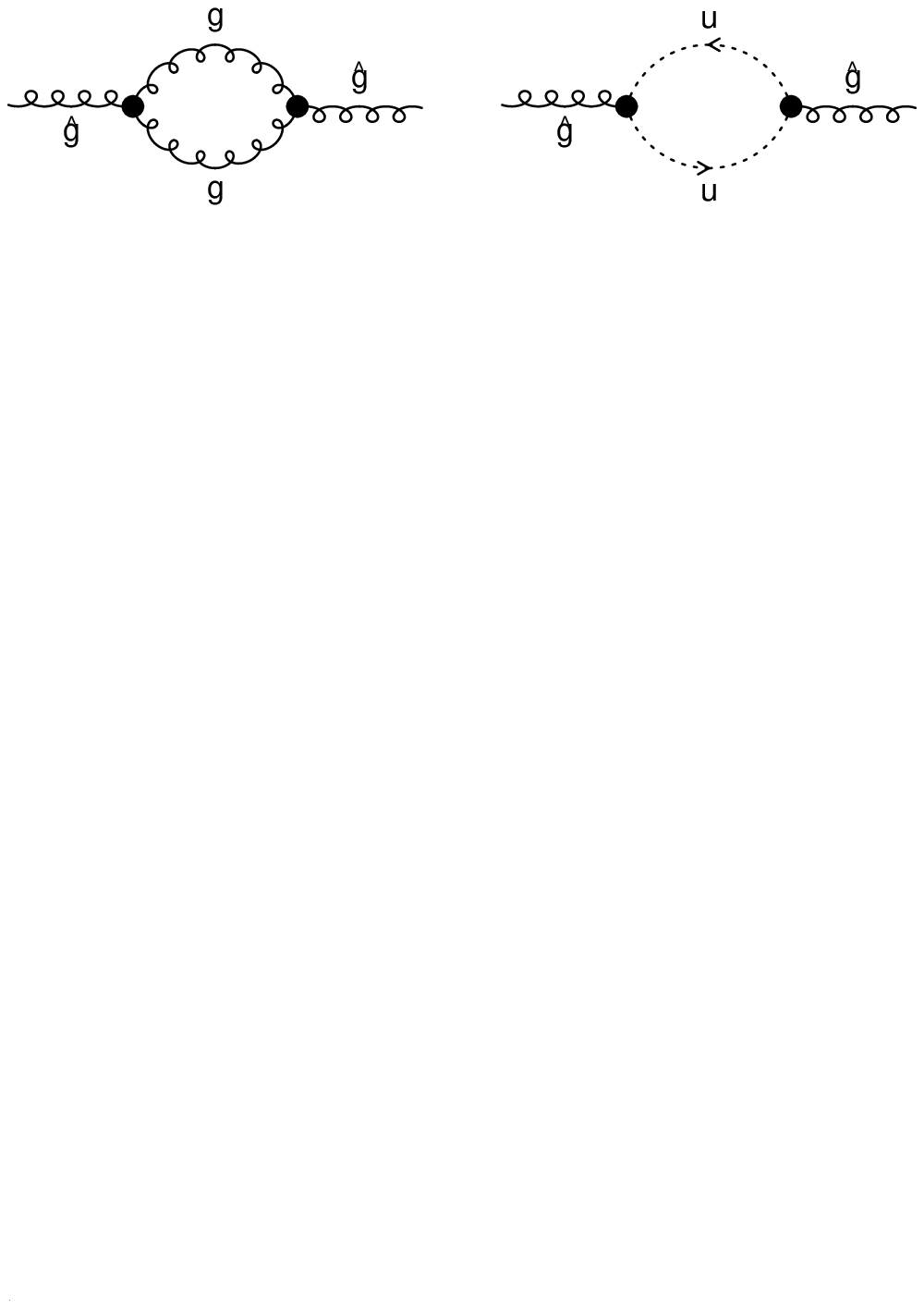}}
\end{picture}
\caption{Feynman diagrams contributing to the one-loop gluon \se.}
\label{FIgg}
\efi
Calculating $\Sigma^{a b}_{\mu\nu}(q)$ in the \bfm\ for arbitrary $\xiQ$
yields
\beq 
\Si^{ab}_{\mu\nu}(q) = -\de^{ab}(q^2g_{\mu\nu} - q_\mu q_\nu) \Bigl[b g^2
\log\frac{|q^2|}{\mu^2} + c(\xiQ) \Bigr],
\eeq
where 
\beq
b=\frac{11N}{48\pi^2},
\eeq
and $c(\xiQ)$ are terms independent of $\mu$ and $q^2$. The coefficient
$b$ of the renormalization-scale-dependent logarithm does not depend on
$\xiQ$ and coincides with the coefficient of $-g^3$ in the $\be$
function $\be(g)$. This is a direct consequence of the gauge invariance
in the \bfm\ \cite{Ab81}.

For the one-loop contribution to the three-gluon vertex
one obtains in the Feynman gauge from the diagrams in \reffi{FIggg}
($k_1= k+q_2$, $k_2=k-q_1$, $k_3=k$) 
\beqar \label{ggg}
\lefteqn{\Gamma^{abc,(1)}_{\mu\nu\rho}(q_1,q_2,q_3) = -ig f^{abc}
\frac{ g^2 N}{2(2\pi)^D} \mu^{4-D}\int{\d^D k} \hspace{8cm} } \nl
&& \biggl[\frac{1}{k_1^2k_2^2k_3^2}
[-\GVVV^{\al}_{\phantom{\al}\mu\be}(k_2,q_1,-k_3)
\GVVV^{\be}_{\phantom{\be}\nu\la}(k_3,q_2,-k_1) 
\GVVV^{\la}_{\phantom{\la}\rho\al}(k_1,q_3,-k_2) \nl[1ex]
&&\qquad\qquad
{}+ 2(k_2+k_3)_\mu (k_1+k_3)_\nu (k_1+k_2)_\rho]\nl[2ex]
&&{}-
 \frac{1}{k_2^2k_3^2}8(q_{1\rho} g_{\mu\nu}-q_{1\nu}g_{\mu\rho}) 
-\frac{1}{k_1^2k_3^2}8(q_{2\mu} g_{\rho\nu}-q_{2\rho}g_{\mu\nu})
-\frac{1}{k_1^2k_2^2}8(q_{3\nu} g_{\mu\rho}-q_{3\mu}g_{\nu\rho})\biggr].
\eeqar
Here again the first term
results from the gauge-boson loop (\reffi{FIggg}a) and the second term from 
the two oppositely directed ghost loops (\reffi{FIggg}b). 
The two-point contributions originate directly from the diagrams
represented in \reffi{FIggg}c;
the diagrams of \reffi{FIggg}d vanish.
\bfi
\begin{picture}(350,180)
\put(30,160){a)}
\put(170,160){b)}
\put(30,63){c)}
\put(170,63){d)}
\put(-40,-360){\includegraphics{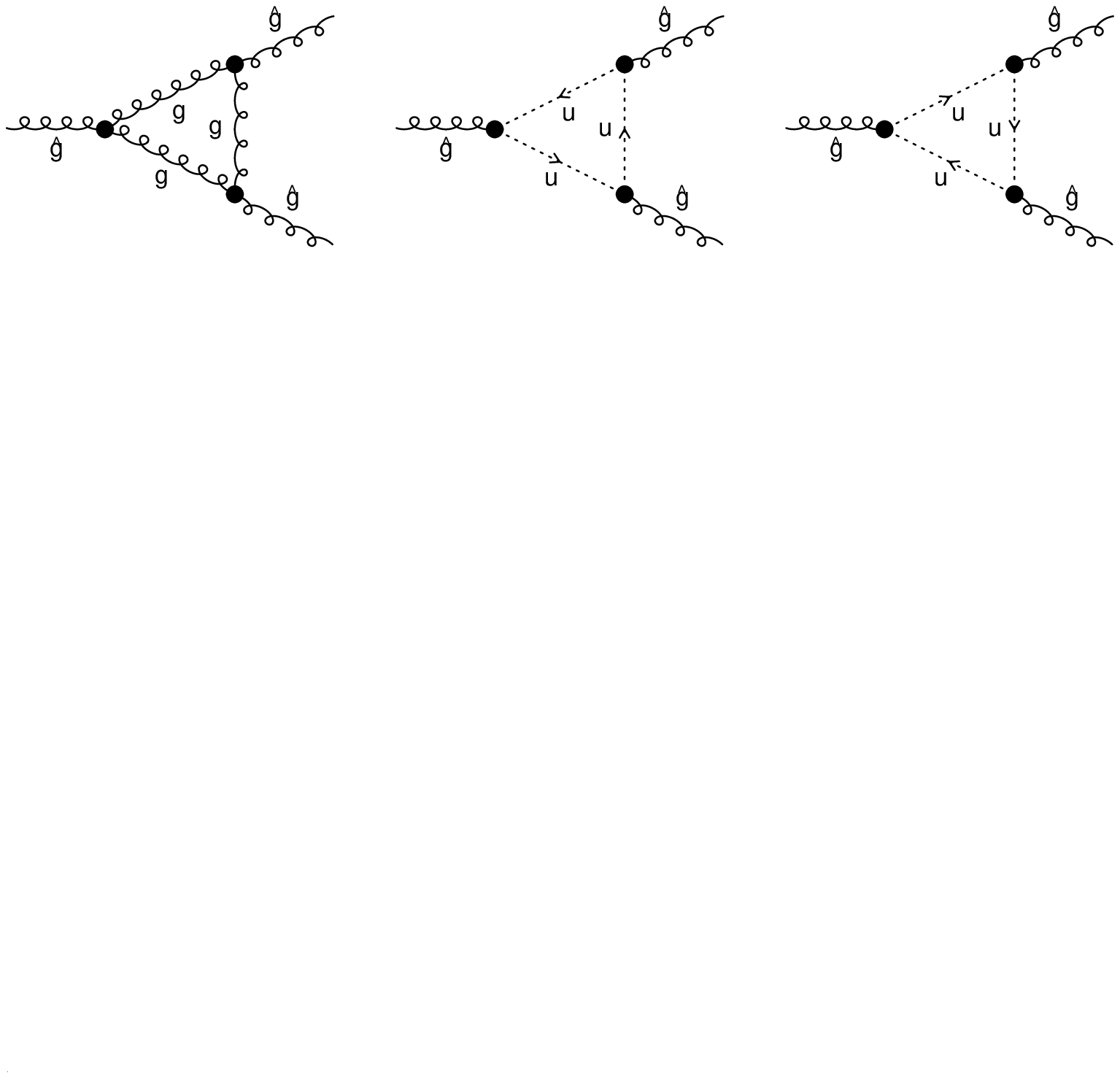}}
\put(-40,-455){\includegraphics{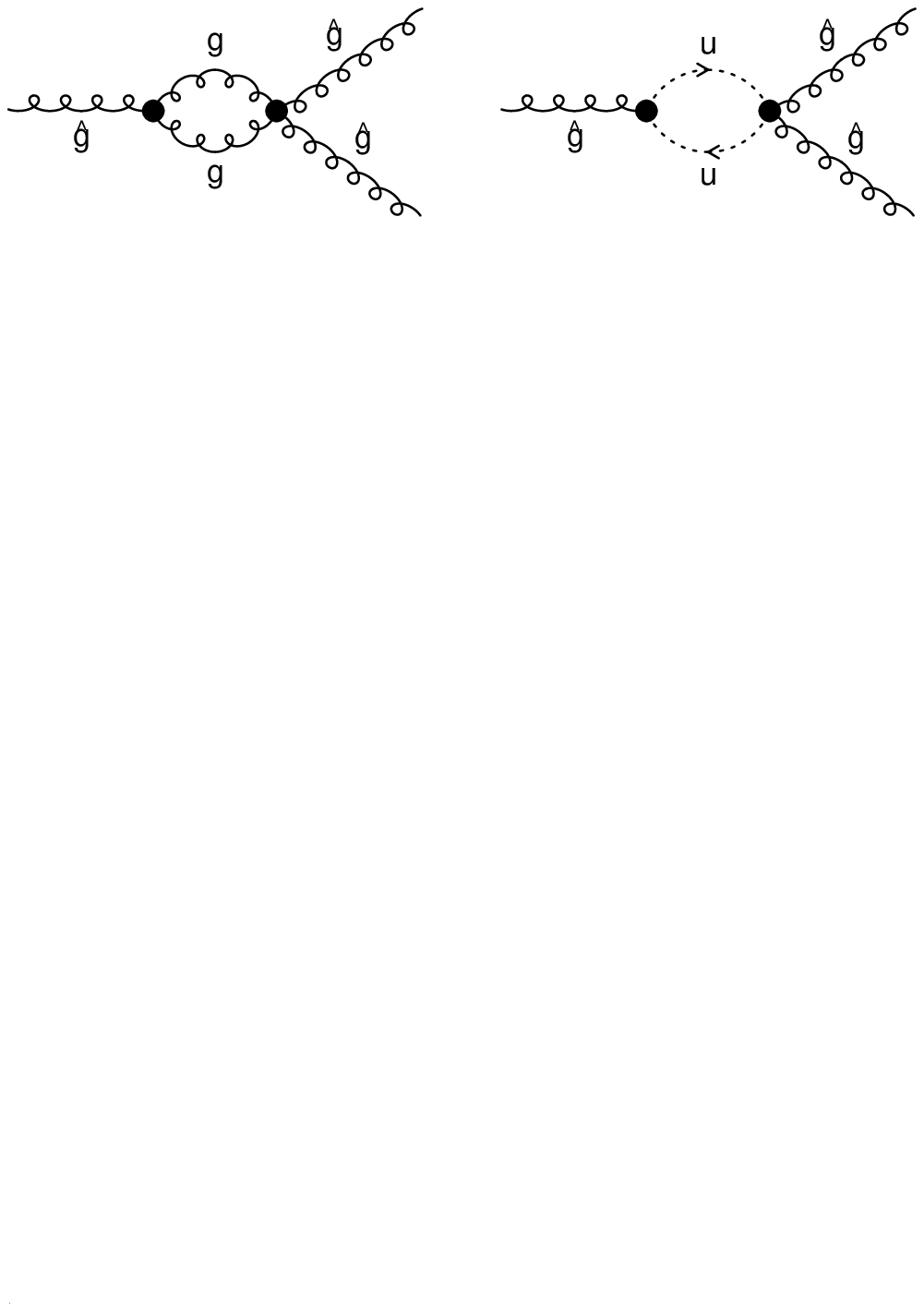}}
\end{picture}
\caption{Feynman diagrams contributing to the three-gluon vertex at the
one-loop level. Each graph in the second line represents three
different permutations. }
\label{FIggg}
\efi
We have applied the \pt\ and checked that it yields exactly the same result.%
\footnote{Note that the signs in \citere{Co89} are inconsistent.}

While in the \pt\ the results \refeq{gg} and \refeqf{ggg} were obtained after
an involved rearrangement of terms between different Green functions, in
the \bfm\ they 
are just the 2-point and 3-point vertex functions, respectively. 
They follow directly from the Feynman rules and are manifestly 
process-independent.

Moreover, the Ward identity which relates the \se\ and the vertex,
\beq
q_{1}^{\mu}\Gamma^{abc,(1)}_{\mu\nu\rho}(q_1,q_2,q_3) = 
-g\left[ f^{abd}\Sigma^{dc}_{\nu\rho}(-q_3)
-f^{adc}\Sigma^{bd}_{\nu\rho}(q_2) \right],
\eeq
holds in the \bfm\ not only for $\xiQ=1$ but 
for arbitrary $\xiQ$. The same is true for the Ward identity relating 
the three- and four-gluon vertices stated in \citere{Pa93}.
While in the \pt\ the validity of these Ward identities was only
verified by explicit computation at one-loop order,
in the \bfm\ they are a direct
consequence of the gauge-invariance of the effective action 
$\Gamma[\Vhat]$ and therefore valid to all orders.
\bigskip

The \bfm\ can be directly applied to the Glashow--Salam--Weinberg
model as well. However, 
in order  to avoid tree-level mixing between the
gauge bosons and the corresponding unphysical Higgs bosons one 
has to generalize the 't~Hooft gauge to the \bfm.
In this case, also the Higgs field
has to be split into a background and a quantum  part. 
While the background Higgs field $\hat\Phi$ has the usual non-vanishing vacuum
expectation value, the one of the quantum Higgs field $\Phi$
is zero. 
Denoting all background fields with a caret, the background-field
't~Hooft gauge-fixing term reads \cite{Sh81}
\beqar\label{tHgf}
\L_{\mathrm{GF}} &=& - \frac{1}{2\xiQ^W}
\biggl[(\de^{ac}\partial_\mu + g_2 \veps^{abc}\hat W^b_\mu)W^{c,\mu}
       -ig_2\xiQ^W\frac{1}{2}(\hat\Phi^\dagger_i \si^a_{ij}\Phi_j 
                  - \Phi^\dagger_i \si^a_{ij}\hat\Phi_j)\biggr]^2 \nl
                 && {}- \frac{1}{2\xiQ^B}
\biggl[\partial_\mu B^{\mu}
       +ig_1\xiQ^B\frac{1}{2}(\hat\Phi^\dagger_i \Phi_i 
                  - \Phi^\dagger_i \hat\Phi_i)\biggr]^2, 
\eeqar
where we have used the conventions of \citere{Dehab}, and $\si^a$,
$a=1,2,3$, denote the Pauli matrices.
We note that 
this gauge-fixing term translates to the conventional one upon
replacing the background Higgs field by its vacuum expectation value and
omitting the background $SU(2)_W$ triplet field $\hat W^a_\mu$.
Background-field gauge invariance restricts the number of 
quantum gauge parameters
to two, one for $SU(2)_W$ and one for $U(1)_Y$.

We note in passing that the \bfm\ has already been applied to the
Glashow--Salam--Weinberg model in \citere{Be93}. However, the gauge-fixing
term used there breaks back\-ground-field gauge invariance as 
no background-Higgs field has been introduced. Since this only concerns 
contributions involving Higgs-particles outside loops, the results of
\citere{Be93} are nevertheless unaffected.

We have evaluated the complete Feynman rules within the 
background-field~'t~Hooft gauge
\refeq{tHgf} including the associated ghost terms for $\xiQ=\xiQ^W=\xiQ^B$. 
The vertices
that do not contain background Higgs fields coincide for $\xiQ=1$ with those of 
\citere{Be93}.
Despite the fact that, owing to the doubling of the fields,
the Feynman rules seem to become more complicated actual
calculations become in fact simpler. This is in particular the case
in the 't~Hooft--Feynman gauge. The number of diagrams contributing
to a certain vertex function is approximately the same as in the
conventional formalism, but the diagrams themselves become easier to 
evaluate. 
Moreover, the number of diagrams contributing to the full (reducible) Green
functions can be reduced by choosing an appropriate background gauge, 
\eg the unitary gauge or a non-linear gauge.

Based on those Feynman rules, we have evaluated the 
quantities that have been treated in the literature using the \pt, \ie
the gauge-boson \ses, the fermion--gauge-boson vertices, and the 
triple-gauge-boson vertices. 
We find that all of them coincide 
for $\xiQ=1$ with those obtained in the \pt~\cite{De92,Pa93b,Pa93a}.
Moreover, all desirable properties known for  the \pt\
vertex functions hold within the \bfm\ for arbitrary $\xiQ$.

The validity of simple Ward identities follows in the
\bfm\ from the manifest gauge invariance. 
We list some of those Ward identities for illustration using the
conventions of \citere{Dehab} (all fields and momenta are incoming and for the 
2-point functions only the momentum of the first field is given):
\beqar
\lefteqn{k^\mu \Ga^{\hat\PA\hat\PA}_{\mu\nu}(k) = 0, \qquad {\mathrm{\ie}} \qquad  
\Si^{\hat\PA\hat\PA}_L(k^2) = 0,} \\
\label{WIAZ}
\lefteqn{k^\mu \Ga^{\hat\PA\hat\PZ}_{\mu\nu}(k) = 0, \qquad {\mathrm{\ie}} \qquad  
\Si^{\hat\PA\hat\PZ}_L(k^2) = 0,} \\
\lefteqn{k^\mu \Ga^{\hat\PA\hat\chi}_{\mu}(k) = 0, \qquad {\mathrm{\ie}} \qquad
\Si^{\hat\PA\hat\chi}(k^2) = 0,} \\
\lefteqn{k^\mu \Ga^{\hat\PZ\hat\PZ}_{\mu\nu}(k) -i\MZ \Ga^{\hat\chi\hat\PZ}_\nu(k) = 0,} \\
\lefteqn{k^\mu \Ga^{\hat\PZ\hat\chi}_{\mu}(k) -i\MZ \Ga^{\hat\chi\hat\chi}(k) 
+\frac{ie}{2\sw\cw} \Ga^{\hat\PH}(0) = 0,} \\
\lefteqn{k^\mu \Ga^{\hat\PW^\pm\hat\PW^\mp}_{\mu\nu}(k) 
\mp\MW \Ga^{\hat\phi^\pm\hat\PW^\mp}_\nu(k) = 0,} \\
\lefteqn{k^\mu \Ga^{\hat\PW^\pm\hat\phi^\mp}_{\mu}(k) 
\mp\MW \Ga^{\hat\phi^\pm\hat\phi^\mp}(k) 
\pm\frac{e}{2\sw} \Ga^{\hat\PH}(0) = 0,} \\[1ex]
\label{WIAff}
\lefteqn{k^\mu \Ga^{\hat\PA\Pfbar\Pf}_{\mu}(k,\bar p,p) = -e\Qf
[\Ga^{\Pfbar\Pf}(\bar p) - \Ga^{\Pfbar\Pf}(-p)],} \\
\label{WIZff}
\lefteqn{k^\mu \Ga^{\hat\PZ\Pfbar\Pf}_{\mu}(k,\bar p,p) -i 
\MZ \Ga^{\hat\chi\Pfbar\Pf}(k,\bar p,p)= e
[\Ga^{\Pfbar\Pf}(\bar p)(\vf-\af\ga_5) - (\vf+\af\ga_5) \Ga^{\Pfbar\Pf}(-p)],} \\
\label{WIWff}
\lefteqn{k^\mu \Ga^{\hat\PW^\pm\Pfbar'\Pf}_{\mu}(k,\bar p,p) \mp 
\MW \Ga^{\hat\phi^\pm\Pfbar'\Pf}(k,\bar p,p)= \frac{e}{\sqrt{2}\sw}
[\Ga^{\Pfbar'\Pf'}(\bar p)\frac{1-\ga_5}{2} 
- \frac{1+\ga_5}{2} \Ga^{\Pfbar\Pf}(-p)],} \\[2ex]
\label{WIAWW}
\lefteqn{k^\mu \Ga^{\hat\PA\hat\PW^+\hat\PW^-}_{\mu\rho\si}(k,k_+,k_-) = e
[\Ga^{\hat\PW^+\hat\PW^-}_{\rho\si}(k_+) - \Ga^{\hat\PW^+\hat\PW^-}_{\rho\si}(-k_-)],}\\ 
\label{WIWAW}
\lefteqn{k_+^\rho \Ga^{\hat\PA\hat\PW^+\hat\PW^-}_{\mu\rho\si}(k,k_+,k_-) 
-\MW \Ga^{\hat\PA\hat\phi^+\hat\PW^-}_{\mu\si}(k,k_+,k_-) =} \nl* && \qquad \qquad
 e\left[\Ga^{\hat\PW^+\hat\PW^-}_{\mu\si}(-k_-) - \Ga^{\hat\PA\hat\PA}_{\mu\si}(k)
+\frac{\cw}{\sw} \Ga^{\hat\PA\hat\PZ}_{\mu\si}(k)\right], \\ 
\label{WIWWA}
\lefteqn{k_-^\si \Ga^{\hat\PA\hat\PW^+\hat\PW^-}_{\mu\rho\si}(k,k_+,k_-) 
+\MW \Ga^{\hat\PA\hat\PW^+\hat\phi^-}_{\mu\rho}(k,k_+,k_-) =} \nl &&  \qquad \qquad
-e\left[\Ga^{\hat\PW^-\hat\PW^+}_{\mu\rho}(-k_+) - \Ga^{\hat\PA\hat\PA}_{\mu\rho}(k)
+\frac{\cw}{\sw} \Ga^{\hat\PA\hat\PZ}_{\mu\rho}(k)\right]. 
\rule{6.5cm}{0mm}
\eeqar
These identities hold for arbitrary $\xi_Q$ in all orders of 
perturbation theory. Note that
the vertex functions are one-particle irreducible, \ie they contain no
tadpole contributions; these appear explicitly as $\Ga^{\PH}(0)$.
We have checked the validity of all  these Ward identities at
tree level and one-loop level by explicit computation. 
Within the \pt\ they have been verified partially in \citeres{De92,Pa93b,Pa93a}.

Having derived the Ward identities in the \bfm, we can also
explain the other desirable properties of the vertex functions noticed
within the \pt.
To this end, we use these Ward identities, simple power-counting
arguments, and the fact that in contrast to the \pt\ the building blocks in 
the \bfm\ are genuine vertex functions.

We first discuss the list of properties stated by
Degrassi and Sirlin \cite{De92}:
\begin{itemize}
\item[(i)] 
The fact that the pole positions of the propagators are not
modified within the \bfm\ at the one-loop level is a direct consequence of the 
gauge invariance of those poles, which are directly related to physical
observables.
\item[(ii)]
The vanishing of the photon--Z-boson mixing at zero momentum,
$\Si^{\hat\PA\hat\PZ}_T(0) = 0$,
follows from the Ward identity \refeq{WIAZ} and the
analyticity of $\Si^{\hat\PA\hat\PZ}_{\mu\nu}(k)$ at $k^2=0$, in analogy to 
$\Si^{\hat\PA\hat\PA}_T(0) = 0$.
\item[(iii)]
 The equality of the \PZ\ and \PW\ \se\ in the limit
 $\sin\theta_W\to0$ and fixed $g_2$ results from the fact that the global
$SU(2)_W$ symmetry is not broken by the background-field  gauge-fixing term.
\item[(iv)]
The UV finiteness of the fermion--gauge-boson vertex functions including
fermion wave-function renormalization 
can be derived from \refeqs{WIAff}, \refeqf{WIZff} and \refeqf{WIWff} 
together with power-counting arguments like in QED. 
\item[(v)]
As stated in \citere{De92}, the fact that the asymptotic behaviour 
for $|q^2|\to\infty$ of the
running couplings defined directly via Dyson summation of the \ses\
is automatically governed by the renormalization group is a
consequence of the UV finiteness of the fermion--gauge-boson
vertices.
\end{itemize}
Moreover, the IR finiteness of the \pt\ \ses\ is trivial within
the \bfm.

We stress that all these arguments do not only hold for $\xiQ=1$ but for
arbitrary finite values of $\xiQ$. In particular, the coefficients of the
leading logarithms of
the \ses\ turn out to be independent of $\xiQ$ in the high-energy limit. 
We have verified all these properties by explicit calculation of the
relevant one-loop vertex functions for arbitrary
$\xi_Q$.
As a consequence, these features
are not related to the $\xi$ independence of the \pt\ \ses\ as could be
supposed from the \pt\ point of view.

From our results for the $\PA\Pt\Ptbar$ and
$\PZ\Pt\Ptbar$ vertices within the \bfm\ we have obtained
the magnetic dipole moment form factor (MDM) of the top quark.
It coincides with the one obtained within the conventional $R_\xi$-gauge
formalism.
In contrast to the statements in
\citere{Pa93b} but in agreement with \citere{Fu72}, we find
that the MDM vanishes in the limit $|q^2|\to\infty$ for all $\xiQ$. 
This has been checked both numerically
and analytically. Moreover, it can be inferred from  
a simple  power-counting argument for all renormalizable gauges.

In \citere{Pa93a} the three-gauge-boson vertices were derived within the
\pt\ for \PW~bosons coupled to conserved currents and the Ward identity
\refeq{WIAWW} was verified for this case.
These results cannot be used as suitable building blocks in processes where 
this restriction does not apply as \eg in $\ga\ga\to\PWp\PWm$. 
Projecting our general \bfm\ results to the case of conserved currents,
we find agreement with (3.16) of \citere{Pa93a}.%
\footnote{There are some sign errors in Ref.~\cite{Pa93a}.}
For our general off-shell result we have explicitly checked the Ward
identity \refeq{WIAWW} and the other two Ward identities
\refeq{WIWAW} and \refeqf{WIWWA} which have not been
mentioned in the \pt\ context. In contrast to the claim made in
\citere{Pa93a}, we found that the $\PA\PW\PW$ vertex within the \pt\ and
the \bfm\ is not IR-finite. 
However, the anomalous-magnetic-moment form factor and the electric-quadrupole 
moment-form factor are 
IR-finite and vanish  in the limit  $|q^2|\to\infty$.
These last two facts can again be derived in the \bfm\ from the Ward identities 
and power-counting arguments.

In the \pt\ formalism the vertex functions are by construction
independent of the gauge parameters $\xi$ within $R_\xi$ gauges.
The gauge parameters only appear
in the tree propagators joining the vertex functions. In
the \bfm\ formalism these gauge parameters correspond to the
background gauge parameters $\xiA$. In \citere{Pa90} the \pt\ has been
applied to four-fermion processes in a gauge theory with spontaneous
symmetry breaking. It turned out that the \se\ contributions to
these processes could be formulated in different ways. Within the
\bfm\ these different possibilities correspond just to different
background gauges. The representation (4.11) of \citere{Pa90}
corresponds to the unitary gauge ($\xiA\to\infty$),
whereas the representation (4.21) corresponds to the Landau gauge
($\xiA=0$).
Thus the results of \citere{Pa90} follow naturally within the \bfm.

\bigskip

In conclusion, we have shown that the BFM provides a systematic
way --- via direct application of Feynman rules --- 
to obtain Green functions that are derived from a gauge-invariant 
effective action, fulfil simple Ward identities and
in comparison to their $R_{\xi}$-gauge counterparts possess very
desirable properties such as improved IR und UV properties and a
decent high-energy behaviour.

We applied the BFM to those cases for which the PT has been used in
the literature. We found that the PT results coincide with the 
special case of the BFM results with quantum gauge
parameter $\xi_Q = 1$. In contrast to the PT, the BFM can directly
be applied to all orders of perturbation theory 
and all types of vertex functions.
Whereas in the BFM the Ward identities are a strict consequence of
gauge invariance, no general proof of their validity exists in the PT,
and the Ward identities can only be verified in specific examples. Moreover,
the calculation of the vertex functions is much simpler in the 
\bfm\ than in the \pt. While the rearrangement of different 
contributions in the \pt\ is quite cumbersome and not clear for
complicated vertex functions, the calculation within the \bfm\
is comparable or even simpler than the evaluation of the
vertex functions in the conventional formalism.
 
In addition, we have found in all cases considered that
the BFM yields well-behaved vertex functions for
arbitrary values of $\xi_Q$.
This means that the requirement of gauge-parame\-ter
independence used in the PT and former treatments is not the
criterion leading to well-behaved vertex functions.
We showed instead that the desirable properties of the background-field
vertex functions are a direct consequence of the BFM Ward identities.
They reflect the underlying gauge invariance and are manifestly valid
for all values of the quantum gauge parameter $\xiQ$.

Our results show in particular that the choice $\xi_Q = 1$
corresponding to the PT is not distinguished on physical grounds but 
it is only one of arbitrarily many equivalent
possibilities. Of course, the background-field 't~Hooft--Feynman
gauge technically facilitates the calculations. The ambiguity of
the vertex functions --- quantified in the BFM by $\xiQ$ --- is
also inherent in the PT and all other constructions, since the
definite prescription to eliminate the gauge-parameter dependence
appears to be just a matter of convention.

One benefit of the BFM is to make this ambiguity apparent. It
shows that the PT vertex functions
cannot carelessly be used to get
physical predictions. Instead, we propose to use the Ward
identities of the BFM to investigate for which cases physically
meaningful results can be obtained. 

\vskip 2ex plus 2ex minus 2ex
\section*{Acknowledgement}
\vskip -1ex plus 1ex
We thank M.~B\"ohm for useful discussions and for 
reading the manuscript.

\end{document}